# A High-Performance Energy Management System based on Evolving Graph

Guangyi Liu, *Senior Member, IEEE*, Chen Yuan, *Member, IEEE*, Xi Chen, *Senior Member, IEEE*, Jingjin Wu, Renchang Dai, *Senior Member, IEEE*, Zhiwei Wang, *Senior Member, IEEE*

***Abstract*--As the fast growth and large integration of distributed generation, renewable energy resource, energy storage system and load response, the modern power system operation becomes much more complicated with increasing uncertainties and frequent changes. Increased operation risks are introduced to the existing commercial Energy Management System (EMS), due to its limited computational capability. In this paper, a high-performance EMS analysis framework based on the evolving graph is developed. A power grid is first modeled as an evolving graph and then the power system dynamic analysis applications, like network topology processing (NTP), state estimation (SE), power flow (PF), and contingency analysis (CA), are efficiently implemented on the system evolving graph to build a high-performance EMS analysis framework. Its computation performance is field tested using a 2749-bus power system in Sichuan, China. The results illustrate that the proposed EMS remarkably speeds up the computation performance and reaches the goal of real-time power system analysis.**

## I. INTRODUCTION

THE grid operating conditions have changed dramatically due to the increasing penetrations of renewable energy, distributed generation, demand response, and energy storage systems at both power transmission and distribution levels [1], [2]. These integrated components are mostly power electronic interfaced and introduced non-linear characteristics into the existing power grid. The exponentially increasing of the system complexity and the frequent and rapid change of system states bring great challenges to the system operators.

An EMS provides operators capability of monitoring, controlling and optimizing power grid operation. Based on the signal from Supervisory Control And Data Acquisition (SCADA), the functions of EMS are essential to the system operators, e.g. NTP presents a system bus-branch model by analyzing the connectivity component in the system node-breaker model, SE estimates system states based on SCADA measurements, PF provides system operation analysis and "N-1" CA evaluates system security and reliability in contingency scenarios with one component loss. However, in current practice, due to the limited computational capability, the commercial EMS only runs every 1–5 minutes for large power systems, leading to the delay from the real system states and lacking the capability of timely following system state changes within seconds or sub-seconds. If a severe event happens, because of extreme weather or overloading, there may be a large difference between the estimated system states and the true real-time system states. Then it is very difficult for system operators to identify the problem and secure the system operation in a timely manner, probably causing cascading failure and large-scale power outage [3], [4]. To build more robust and reliable power systems, an advanced EMS for system states real-time monitoring and fast response is a necessity [5], [6]. Parallel computing is one promising method to improve computation efficiency by taking advantages of advanced computation technique, rich storage space and parallel capability of processing units. Besides, with the evolution of database, the graph data structure is becoming popular, because of its features of system natural expression, lightweight, low I/O cost, parallelism and quick search.

A graph intuitively represents a system with vertices and edges, and respectively stores system information as attributes of vertices and edges. High-performance graph computing applications in power system analysis have been developed and verified in [7]. To further improve the EMS performance, the technique of evolving graph is a promising and feasible solution [8]. It improves the storage, the communication, and the computation performance by taking advantage of system information in the previous graph snapshots.

This paper presents a high-performance EMS framework based on evolving graph. Evolving graph is able to mimic power systems real-time dynamic variations, including changes in system topology and system states. Besides, the evolving graph based power system applications, like NTP, SE, and "N-1" CA, are more efficient using the power system status from the previous snapshot, since the power system operation is a dynamic continuous process. Furthermore, in CA, each scenario is viewed as a spatial evolving graph of the base case. This is because each CA case is a potential dynamic topology change with one component failure from the base case. Then the time consumption in CA can be significantly reduced by using the base case results for each evolving scenario. Field testing in the Sichuan grid is conducted to explore the performance in computation speed. The field testing results illustrate that the proposed approach is able to significantly improve the performance of the power system analysis in EMS.

The rest of this paper is organized as: Section II introduces

This work is supported by the State Grid Corporation of China technology project 5455HJ180020. (*Corresponding author: Xi Chen*).

G. Liu, C. Yuan, X. Chen, J. Wu, R. Dai and Z. Wang are with GEIRI North America, San Jose, CA 95134 USA (e-mail: {guangyi.liu, chen.yuan, xi.chen, jingjin.wu, renchang.dai, zhiwei.wang}@geirina.net).

graph computing and evolving graph. Power system graph modeling is presented in Section III. Section IV illustrates the EMS framework over evolving graph. The testing is presented in Section V and the paper is concluded in Section VI.

## II. Graph Computing and Evolving Graph

### A. Graph Database

Graph database employs semantic queries with vertices, edges and attributes to store data in a graph structure. A graph is expressed as an ordered pair of vertices and edges, $G = (V, E)$. In the graph model, each vertex, denoted as $V$, represents an entity, and the relationships between these entities are represented by edges, represented by $E$.

### B. Graph Computing

A key feature of a graph database is giving priority to relationships. For any vertex, its neighboring vertices and their attributes can be accessed through its connected edges. Then, the operations can be performed through direct links, thus avoiding the effort to use foreign keys in a relational database. Also, in graph database, a vertex is a storage and computing unit. Local computation is implemented on each vertex, and all vertices can complete local computation independently and in parallel. In Fig. 1, it clearly shows that the graph computing parallelism is obtained with the Bulk Synchronous Parallel (BSP) programming model. In the model, the computation contains super-steps, and, in each super-step, a set of processors perform their local computation independently and simultaneously. After local computation, the processors exchange data by passing message via the communication layer. All the processors synchronize at the end of each super-step. If a processor finishes its local task and reaches the synchronization barrier, it waits for other processors. The parallelism is performed by walking the graph concurrently and computing simultaneously over vertices and edges.

### C. Evolving Graph

The evolving graph is a sequence of graph snapshots in the timescale. It models a dynamically changing system, like social network and power system. Each graph snapshot in the evolving graph represents the modeled system at the corresponding time point, including its topology information, system states, etc. Different from the commonly used graph model [9], which is expressed as a 2-element ordered pair, the $G = (V, E)$ evolving graph is defined as a 3-element tuple $EG = (V, E, t)$, where $EG$ indicates the evolving graph model, $t$ is the time point, $V$ denotes a set of vertices at time $t$, and $E$ represents a set of vertices and edges at time $t$.

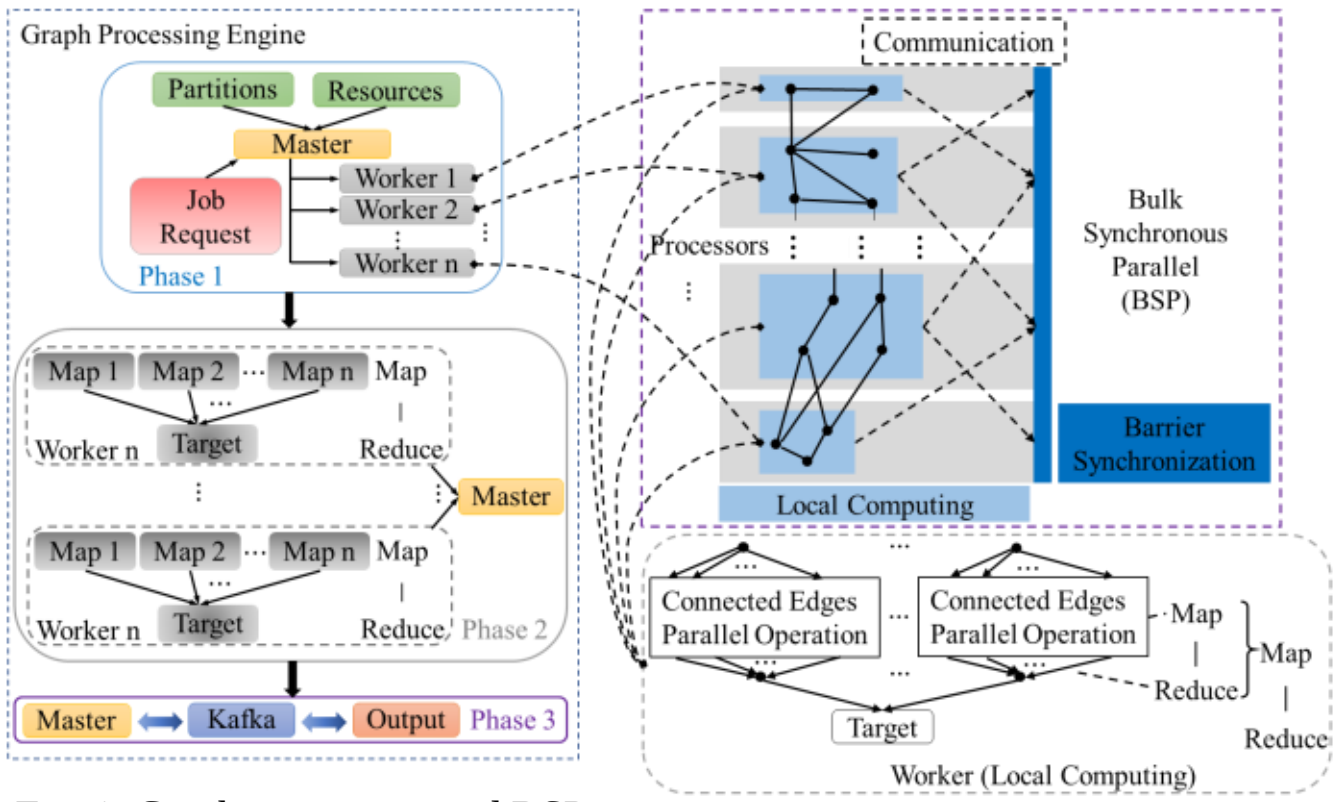

Fig. 1. Graph computing and BSP

An example of the evolving graph is shown in Fig. 2. The graph at time point $t_0$, denoted by $G(t_0)$, is viewed as the base graph. The graphs that follow on time axis are similarly named as $G(t_1), G(t_2), \cdots, G(t_n)$. As time goes, the system network and states change gradually. From $t_0$ to $t_1$, two lines are added, as shown in red, and three nodes have power injection changes, highlighted in purple. At $t_n$, there is an edge disconnection, displayed with a black cross, and also a line reconnection happens, as shown in red. In addition, from $t_{n-1}$ to $t_n$, the system has power injection changes at four nodes.

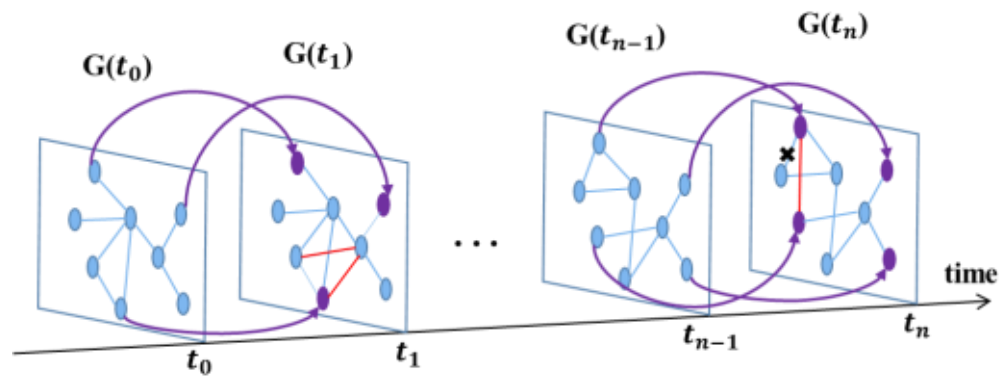

Fig. 2. An example of evolving graph

For the power system network, the intuition is that in most steady-state cases, system topology and status change gradually. If the sampling time is small enough, the system status of two consecutive time sections is expected to be close. In other words, if there is only a minor change in the two continuous system snapshots, system states in a previous time section could be used as the initial start, and components, like bus ordering, matrix formulation, and matrix factorization, in the previous graph could be reused to reduce the processing time spent on power system analysis.

## III. Power System Graph Modeling

Power system network analysis models are categorized into two types: (1) CIM/E based node-breaker model, and (2) bus-branch model. The first model represents the real, physical power grid and illustrates the switching status of circuit breakers and disconnectors for each substation, demonstrating power system device connection within each substation, substation interconnection and network topology. Different from the system node-breaker model, the bus-branch model is an abstract model for system analysis, derived from the node-breaker model. It is usually used for the online and offline power system analysis, such as SE, power flow, CA, etc. The NTP is used to convert the node-breaker model into the bus-branch model.

### A. CIM/E based Node-Breaker Graph Model

Based on common information model (CIM), CIM/E was designed to represent power system components and is an open standard for data exchange between EMS vendors, with a smaller size and simplified form [10]. In Fig. 3(a), the one-line diagram of a substation node-breaker model is illustrated, consisting of bus bars, circuit breakers, disconnectors, loads and generators. A CIM/E representation of this substation is also presented in Fig. 3(a). In Fig. 3(b), a CIM/E based substation is modeled using a graph database [10]. It uses vertices to model all devices, giving operators a straightforward visualization of substation topology and provide them easy access to conduct data management and graph model manipulation.

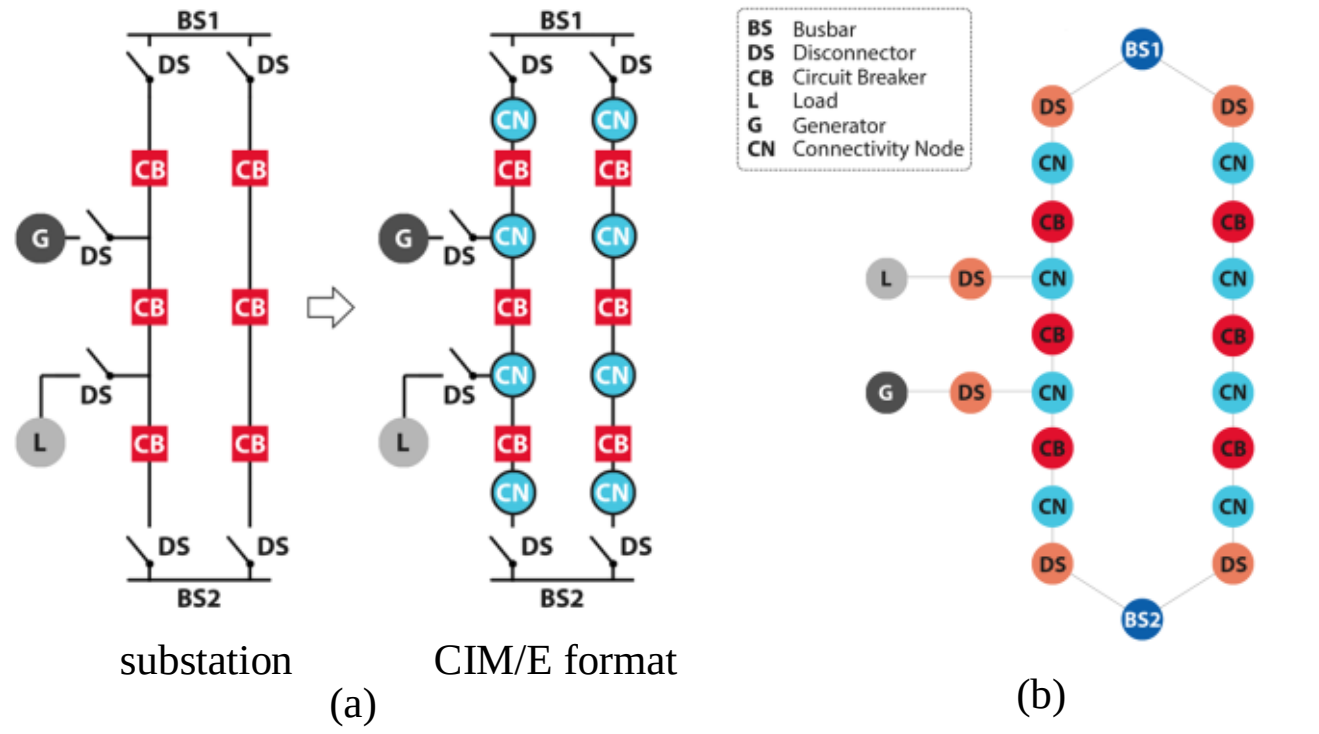

Fig. 3. Substation Node-Breaker model representation in (a) one-line diagram and CIM/E format, and (b) CIM/E based substation graph modeling

## B. Bus-Branch Graph Model – Admittance Graph

The system bus-branch model is developed from system node-breaker model via network topology processing. In the bus-branch graph model, a bus is represented by a vertex, a branch is expressed as an edge, and bus and branch parameters, such as load demand, power generation, bus voltage, including magnitude and phase angle, line resistance, and line reactance, are represented by attributes associated with vertices and edges. Since the bus-branch graph model has the same topology as the system admittance matrix, it is also called admittance graph.

## C. Power System Computing Model – Factor Graph

Most power system analysis is approximately equivalent to solving linear equations in mathematics. Its generalized format is presented in (1).

$$A \cdot x = b. \tag{1}$$

Solving a linear equation with direct method, through LU factorization and forward and backward substitution (F/B substitution), is implemented as a sequence of graph operations in the graph model. It mainly includes three steps, as displayed in Fig. 4.

*(a) Graph Structure Analysis:* $G(A_t)$ represents the graph of matrix $A_t$, where $A_t$ is the matrix $A$ at time $t$. Structure analysis exploits the vertex and edge distribution of $G(A_t)$, which is a non-zero pattern of $A_t$, to find a reordering of the vertices in the corresponding graph, so that the number of fill-ins can be reduced during matrix factorization. Meanwhile, a spanning tree $T(A_t)$, also known as the elimination tree, is created to provide a guidance for factorization. The potential parallelism is indicated by the elimination tree.

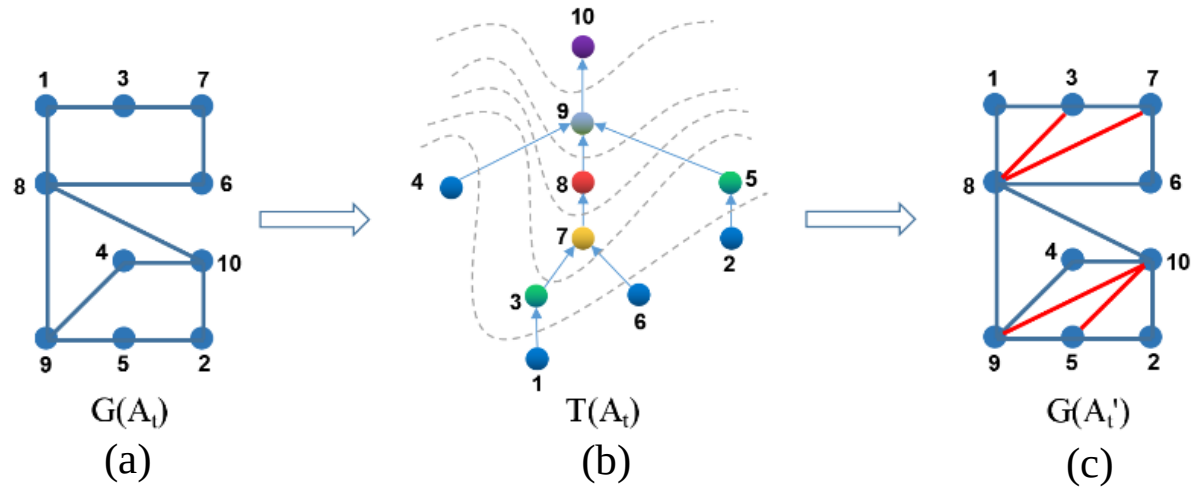

Fig. 4. Graph LU factorization (a) graph of matrix $A_t$, (b) elimination tree, (c) LU graph

*(b) Dynamic Transform on Graph:* Dynamic transform on the graph corresponds to the LU factorization step which factorizes the matrix $A_t$ into the product of a lower triangular matrix, $L_t$, and an upper triangular matrix, $U_t$, i.e., $A_t = L_t U_t$. In the elimination tree $T(A_t)$ in Fig. 4, the vertices with the same color are grouped in the same level and can be computed simultaneously due to their independence to each other. In the graph model, after factorization, the information of $L_t$ and $U_t$ are stored as LU graph $G(A'_t)$ for solving.

*(c) Solving:* Based on the LU graph $G(A'_t)$, the triangular systems $L_t y_t = b_t$ and $U_t x_t = y_t$ are solved via forward and backward substitution respectively.

# IV. Evolving Graph based Power System EMS

As high penetrations of power electronic devices being introduced into the grid, the nonlinear and fast time-varying characteristics are becoming more prominent in the power system. A high-performance EMS is urgently needed to assist system operators and ensure a secure and reliable power system operation. This paper proposes to use evolving graph, including temporal evolving graph and spatial evolving graph, to improve the performance of power system analysis.

## A. NTP over Temporal Evolving Graph

There are two types of network topology processing: (a) full NTP, in which the bus-branch graph model is created from scratch; (b) incremental NTP, in which the breaker status that has dynamically changed since the last snapshot is identified and the NTP is processed only for the status-changed switching devices, considered as dynamic NTP. With the second approach, only the dynamically changed area is rebuilt.

Since full NTP starts from scratch, it needs to process the system topology substation by substation and component by component, consuming large amounts of time. However, in practice, between two consecutive system snapshots, very few circuit breakers and disconnectors change status. Therefore, only at the initial stage, it needs full NTP to build system bus-branch graph model from the system node-breaker graph model, then the incremental NTP is employed to dynamically update the system bus-branch graph model and dramatically improve the NTP performance. It is based on the temporal evolving graph, which stores system historical information and displays system dynamic evolution.

## B. SE over Temporal Evolving Graph

To realize a high-performance power system analysis, SE is the fundamental function and needs to be completed fast, e.g. in sub-seconds, and accurately to well serve its following applications. At each snapshot $t$, the main step of SE is to iteratively solve (2):

$$G(x_t) \cdot \Delta x_t = H^T(x_t) \cdot R^{-1} \cdot (z(t) - h(x_t)), \tag{2}$$

where $x_t$ is the system states vector at time point $t$, $z(t)$ is the system measurement vector at time $t$, $h(x_t)$ is the calculated vector relating $x_t$ to the error free measurements, $H(x_t) = \frac{\partial h(x_t)}{\partial x_t}$ is the Jacobian matrix, $R^{-1}$ is the weight matrix, and $G(x_t)$ is the gain matrix, which is a constant in the fast decoupled mode. Graph computing based fast decoupled state estimation method is derived and well elaborated in the previous work [11]. It decomposed the formulation of system-level state estimation problem into node-based problems in the system graph, and largely reduced its computation time with node-based graph computing for problem formulation and hierarchical graph computing for problem solving. Continuing from that point, if the system graph model does not change, or changes a little, $G(x_{t-1})$ and its corresponding triangular factors $L(x_{t-1})$, $U(x_{t-1})$ developed at last time point can be reused or dynamically updated at time *t*, saving the time spent on gain matrix formulation and factorization. Furthermore, considering the dynamic and continuous time-varying feature of power system operation, system states have little difference between two consecutive snapshots at SCADA rate (5 seconds). Then, with the employment of temporal evolving graph, using system states calculated in the previous snapshot as the initial values could provide a good start point, reduce computing iteration and further save processing time.

### C. CA over Spatial Evolving Graph

CA is a “what if” scenario analysis. It evaluates the impacts on power system operation when potential outages occur in the next time step. From the view of evolving graph, if system bus-branch graph model is viewed as the base case, the contingency related bus-branch graph models form a series of dynamic spatial evolving graphs derived from the base graph. Base case system states are the initial start of each contingency case. Besides, the calculation results such as the bus ordering and matrices, like $Y_{bus}$, $L$ and $U$, stored in the base graph can be efficiently reused. Two approaches are investigated in this paper: 1) graph computing based fast-decoupled power flow [12] (GC-FDPF) by using base case system states as the initial start, and reusing base case bus ordering and symbolic analysis in each contingency scenario; 2) modified preconditioning conjugate gradient by using base case system states as the initial start, reusing base case LU matrices, i.e. $L'_{basecase}U'_{basecase}\Delta\theta_1 = \Delta P_{CA}/|V_{basecase}|$, and employing preconditioning conjugate gradient method to solve $B'_{CA}\Delta\theta_2 = \Delta P_{CA}/|V_{basecase}|$. The second method assumes that system states in “N-1” contingency cases are close to the base case. Then, the convergence and the computation efficiency of the second approach is competitive. The details of this method are well elaborated and verified in the previous study [13].

### D. Evolving Graph based High-Performance Power System EMS Framework

Fig. 5 illustrates the evolving graph based high-performance power system EMS framework. It mainly includes eight steps and dynamically conducts system analysis with SCADA signals every 5 seconds. Step one is the SCADA signal input stage. If the switching status does not change from the previous snapshot, then no topology processing is needed except for system measurements update in step 2 and step 3. Otherwise, the CIM/E based node-breaker graph model is dynamically updated with switching status change in step 2 and the bus-branch graph model is developed via incremental NTP in step 3. After that, based on dynamic and continuous time-varying characteristics of power system operation, the temporal evolving graph is employed for efficient state estimation in step 4 with small dynamics, as described in Section IV.B. Then step 5 returns the estimated states to update the bus-branch graph model and be the initial start for the power flow analysis in step 6. After the power flow calculation, step 7 sends the line power flow and system violation results back to update the bus-branch graph model. Then spatial evolving graph based “N-1” contingency analysis is implemented, as elaborated in Section IV.C.

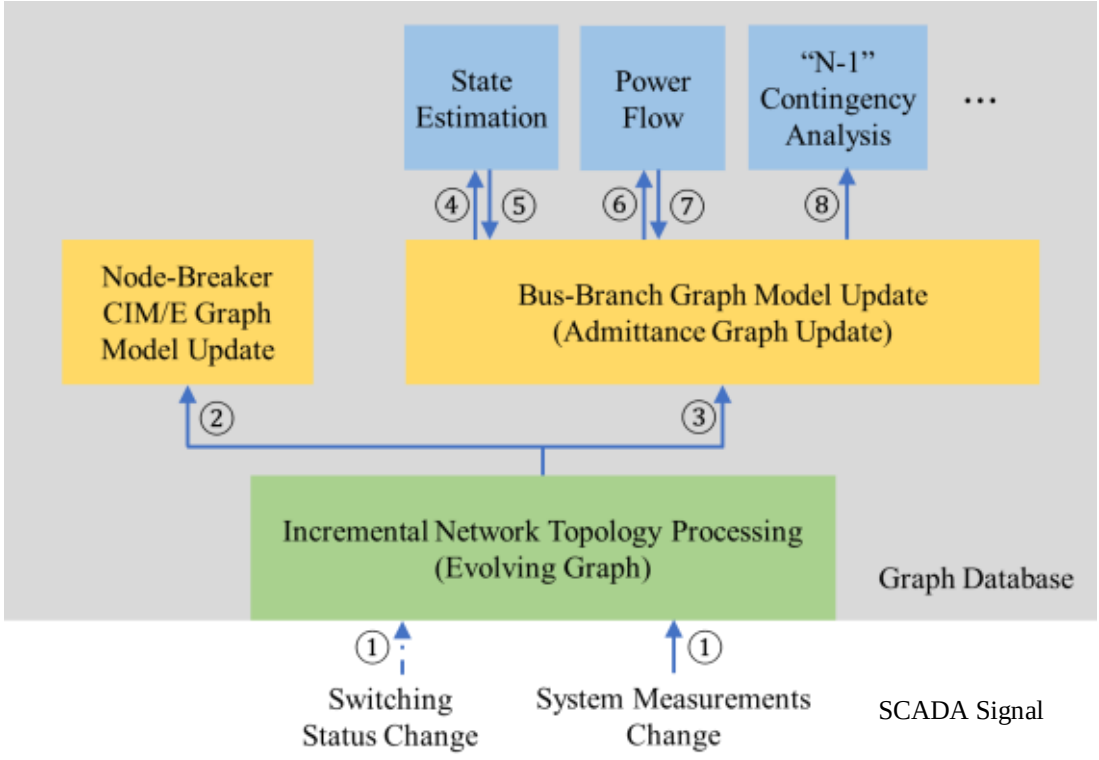

Fig. 5. Evolving graph based high-performance power system EMS framework

## V. Testing Results

The field testing is presented to demonstrate the high performance of the proposed evolving graph based power system EMS. It uses a provincial power system in Sichuan, China, having 2749 buses and 3282 branches. The testing environment in the control center of Sichuan power grid is listed in Table I.

TABLE I. Testing Environment

| **Hardware Environment** | |
|---|---|
| CPU | 2 CPUs × 6 Cores × 2 Threads @ 2.10 GHz |
| Memory | 64 GB |
| **Software Environment** | |
| Operating System | CentOS 6.8 |
| Graph Database | TigerGraph v0.8.1 |

### A. NTP Performance over Temporal Evolving Graph

The execution time for the full NTP and incremental NTP with a few switching status changes are shown in Table II. It is clearly shown that if the NTP is implemented based on an evolving graph, the time consumption is largely reduced.

TABLE II. Network Topology Processing over Temporal Evolving Graph

| Number of Vertices | Number of Edges | Full NTP (ms) | Incremental NTP (ms) (Switching Status Change) |
|---|---|---|---|
| 78411 | 75277 | 600 | 100 |

### B. SE Performance over Temporal Evolving Graph

Using an evolving graph based SE, the time cost is greatly saved. As presented in Table III, if the topology changes, using previous snapshot’s system states as the initial start reduces the number of iterations, even though the gain matrix formulation and LU factorization time still exist. If no topology change, state estimation is quickly implemented by reusing last SCADA snapshot’s system states and gain LU matrices, greatly saving computation time. In this case, only 1 iteration is needed, and the total time is only 7.68 ms.

TABLE III. State Estimation over Temporal Evolving Graph

| Scenario | | Topology Change | No Topology Change |
|---|---|---|---|
| Total Time (ms) | | 54.98 | 7.68 |
| Gain Matrix (ms) | Formulation | 19.48 | — |
| | LU Factorization | 13.33 | — |
| Number of Iterations | | 3 | 1 |
| Per Iteration (ms) | RHS Vector Update | 6.06 | 6.08 |
| | F/B Substitution | 0.33 | 0.30 |
| | System States Update | 0.96 | 0.98 |

### C. *CA Performance over Spatial Evolving Graph*

*(a) Scheme 1: Using GC-FDPF:* GC-FDPF computation time analysis for Sichuan power grid is displayed in Table IV. Based on the evolving graph, CA scenarios are evolved from the base case. For each scenario, the computation time in symbolic analysis, which is over 30% of the total time cost on each contingency, is saved. In addition, using base case system states as the initial start, the number of iterations in power flow calculation would be largely reduced, and the initialization time is almost saved since the base case did the initialization and stored the information in the graph model.

TABLE IV. GC-FDPF Time Analysis

| Initialization (ms) | Solve $B' \cdot \Delta\theta = \Delta P/V$ , $B'' \cdot \Delta V = \Delta Q/V$ | | |
|---|---|---|---|
| | Symbolic Analysis (ms) | Numerical Factorization (ms) | Solve (ms) |
| 6.32 | 6.94 | 0.67 | 7.71 |

*(b) Scheme 2: Using Graph Computing based Modified Preconditioning Conjugate Gradient:* Different from fast-decoupled power flow, the modified preconditioning conjugate gradient method's convergence highly depends on the specific line outage [13]. Non-critical line outage needs few iterations, while a critical line-outage costs more. Based on the difference of individual case, 514 scenarios are analyzed, ignoring islanding and end-point isolation scenarios. This is because the isolated parts in cases of islanding and end-point isolation introduce divergent issues to the system analysis process. The total computation time is only 60 ms.

### D. *Performance Comparison with Commercial EMS*

This section provides a computation performance comparison between the proposed evolving graph based EMS prototype and a state-of-the-art commercial EMS – D5000 system operating in the control center of Sichuan power grid in China. The proposed EMS system integrates the verified NTP, SE, and CA applications. Besides, for the effectiveness and validity of the performance assessment, the proposed EMS is installed and tested in the control center of Sichuan power grid. Furthermore, to ensure the two EMS systems using the same snapshot for the performance comparison, the communication between the proposed EMS system and the SCADA system was built up through socket to receive dynamic SCADA measurement and switching signals in real-time, the same as the commercial EMS in Sichuan power grid. The computation performance of both systems is evaluated in Fig. 6. It clearly demonstrates that the period of the proposed evolving graph based EMS is within a SCADA sampling rate, 4-5 seconds, and its computation performance is much better than the available EMS in the commercial market.

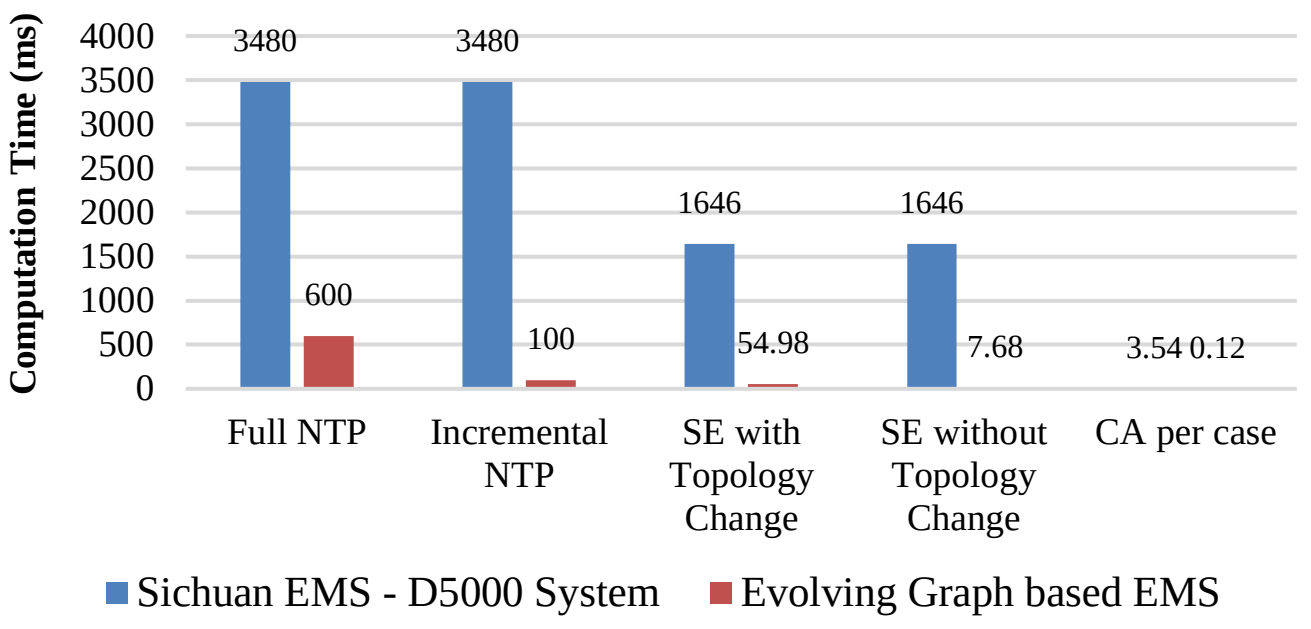

Fig. 6. EMS computation performance comparison in Sichuan power grid

## VI. Conclusion

In this paper, a high-performance EMS using the evolving graph is proposed. Graph modeling in power systems, like node-breaker graph model, bus-branch graph model, and factor graph model, are developed. Based on these graph models, power system EMS is built over evolving graph for high-performance NTP, SE, "N-1" CA, and other advanced EMS functions. The field testing of a 2749-bus system in Sichuan province, China, indicates the dramatical improvement in the performance of power system analysis by reusing the knowledge of the previous snapshot.